# Reduced Bond Graph via machine learning for nonlinear multiphysics dynamic systems

**Youssef Hammadi [a,b], David Ryckelynck [b,*], Amin El-Bakkali [a]**

[a] *Methods and Tools for Numerical Simulation Department, Renault Group, Guyancourt Technical Center, 1 Avenue du Golf, 78280 Guyancourt, France*

[b] *Simulations of Materials and Structures Department, MINES ParisTech - PSL University, Center of Materials, CNRS UMR 7633, 63-65 Rue Henri Auguste Desbruères, BP 87, 91003 Evry, France*

**ABSTRACT**

We propose a machine learning approach aiming at reducing Bond Graphs. The output of the machine learning is a hybrid modeling that contains a reduced Bond Graph coupled to a simple artificial neural network. The proposed coupling enables knowledge continuity in machine learning. In this paper, a neural network is obtained by a linear calibration procedure. We propose a method that contains two training steps. First, the method selects the components of the original Bond Graph that are kept in the Reduced Bond Graph. Secondly, the method builds an artificial neural network that supplements the reduced Bond Graph. Because the output of the machine learning is a hybrid model, not solely data, it becomes difficult to use a usual Backpropagation Through Time to calibrate the weights of the neural network. So, in a first attempt, a very simple neural network is proposed by following a model reduction approach. We consider the modeling of the automotive cabins thermal behavior. The data used for the training step are obtained via solutions of differential algebraic equations by using a design of experiment. Simple cooling simulations are run during the training step. We show a simulation speed-up when the reduced bond graph is used to simulate the driving cycle of the WLTP vehicles homologation procedure, while preserving accuracy on output variables. The variables of the original Bond Graph are split into a set of primary variables, a set of secondary variables and a set of tertiary variables. The reduced bond graph contains all the primary variables, but none of the tertiary variables. Secondary variables are coupled to primary ones via an artificial neural network. We discuss the extension of this coupling approach to more complex artificial neural networks.

## 1   Introduction

Nowadays, engineering systems are getting more and more large, nonlinear and integrated. The Bond Graph (BG) approach [1] allows the modeling of such systems thanks to their ability to account for multiphysics interacting systems in a unified manner. These bond graph models rely on a system of Differential Algebraic Equations (DAEs), responding to some input excitations and resulting in few outputs useful in engineering. As part of this work, we will use simulation data generated by bond graph models as inputs to Machine Learning (ML) algorithms in order to setup a reduced DAE system (RDAE) related to a reduced Bond Graph (RBG). This work belongs to hybrid strategies that marry Ordinary Differential Equations (ODEs), or DAEs, with machine learning [2,3]. We are using the inherent generative capability of ODEs or DAEs, not for data augmentation as in [2], but for Bond Graph reduction. Although there are promising results on physics-based machine learning [4], it has been underutilized in engineering applications involving the solution of DAEs or ODEs. In this work, the RBG is supplemented by an artificial neural network (ANN). Therefore, the residual of the DAE system is not fully replaced by a recurrent neural network as proposed in [4] when using physics-based machine learning. The output of the proposed machine learning involves a set of DAEs. It is not solely data. Hence, it is difficult to use an usual Backpropagation Through Time [5] for the optimization of the weights involved in the ANN. We can follow the reverse-mode derivative of an ODE initial value problem, proposed in [3]. But it is rather an intrusive approach when using some ODEs industrial solvers where no automatic differentiation is available. So, as a first attempt, a very simple ANN is proposed by following a model reduction approach, via unsupervised machine learning. A surrogate model that balance well accuracy and complexity is called proper [6]. Proper modeling techniques can be based either on model deduction or model order reduction (MOR) [7]. In model reduction approaches we start from a complex model, then reduce the number of its variables, such that it becomes proper. A literature review of proper modeling techniques [7,8] shows that each of the existing methods has one or more of the five following limitations:

1. Applicability to a restricted class of systems: many proper modeling techniques are applicable only to limited sets of systems such as linear systems or time-invariant ones.

2. Requiring a change of the state variables original meaning: some model reduction methods, such as projection-based ones, require a change of states original meaning, thus losing their physical intuitive denotation. In this paper, a proper modeling technique that preserves the original meaning of the states is referred to as *states meaning-preserving*.

3. Input independence: some of the existing methods build proper models that are independent from the input excitations, where the term "input" refers as in [9] to excitations as well as parameters and initial conditions. However, it may be preferable, in many engineering use cases, to design a model which is proper only for a given family of inputs trajectories. In this paper, a proper modeling technique filling this need is referred to as *input-dependent method*.

4. Inapplicability to graph models: some simplification methods are applicable only at the equation level, without a clear reflection at the graph-level representation.

5. Not reducing the bond graph junction structure: many reduction methods focus on reducing the number of states, without considering bond graph structure reduction.

---

[*] Corresponding author.
E-mail address: david.ryckelynck@mines-paristech.fr

Moreover, in our point of view, there is a sixth limitation:

6. Inability to reconstruct the deleted variables: some reduction methods act by eliminating the less significant variables, thus making their usage impossible in the Reduced Order Model (ROM).

Auto-encoders [10,11] have been proposed to both reduce the dimension of data and decode the reduced data to recover the high dimensional data. In [3], a neural decoder is proposed to recover high dimensional state variables by using reduced variables. In this paper, a simpler decoder is proposed, but it enables to reduce the original Bond Graph.

In bond graphs proper modeling literature, we find some model deduction techniques [12,13], and other model reduction ones: mainly modal analysis [14] and energy-based MOR methods [15]. The modal analysis approach expresses the reduced model in terms of modal coordinates instead of physical ones, and is therefore not a states meaning-preserving method. Note that, in its simplest rendition [14], modal analysis method was proposed for linear finite-dimensional systems, before some extensions were made to nonlinear systems [16]. Energy-based methods reduce models by eliminating the studied system less energetic components, while minimizing the effect of these eliminations on the overall energy flow. Three techniques are presented in [15]. Each of the three techniques is based on a particular energy metric, and overcomes the limitations 1 to 5. Energy-based MOR methods were used to reduce several bond graph models as illustrated in [15], [17] and [18].

As part of the present work, we propose a methodology that circumvents the above limitations using unsupervised machine learning applied to simulation data. As a result, we obtain a hybrid model marrying machine learning and physics-based modeling. In practice, a RBG is coupled to an artificial neural network. In order to cover the variation of the model inputs, we generate the simulation data using a Design of Experiment (DOE). We will distinguish the training inputs sets to build the RBG, and the testing inputs sets to validate the already built RBG. Besides addressing the limitations 1 to 5, the obtained RBG surpass the sixth limitation by recovering the eliminated variables thanks to a reconstruction ANN as explained in next section.

In the outline of this paper, we will first present the methodology through a reduction process for nonlinear DAE systems (section 2). Then, we will reinterpret the reduced DAE system in bond graph form through an illustrative example (section 3). Finally, the methodology is applied to an industrial cabin thermal model in order to assess the speedup and the accuracy of the RBG (section 4). In this last section, we will generate the simulation data through a DOE over 7 cabin inputs in order to make the RBG valid for a wide range of model inputs. This allows using the RBG in some simulations with time-variant inputs such as homologation driving cycles (variant vehicle speed) and cabin temperature control loops (variant temperature, flow rate and humidity of cabin inlet air). The paper ends with a Discussion section about the optimization of the neural network.

## 2 Method

### 2.1 *DAEs related to the original Bond Graph*

Any BG model relies, at the equation level, on a DAE system. The variables of the DAE system are denoted by two column vectors $\boldsymbol{\theta}(t,\boldsymbol{\mu}) \in \mathbb{R}^{\mathcal{N}_\theta}$ and $\boldsymbol{\gamma}(t,\boldsymbol{\mu}) \in \mathbb{R}^{\mathcal{N}_\gamma}$, where $t$ is the time variable and $\boldsymbol{\mu} \in \mathbb{R}^{\mathcal{N}_\mu}$ is the vector of BG inputs which can be either excitations or parameters of the BG model. $\boldsymbol{\theta}$ is the vector of differential variables, and $\boldsymbol{\gamma}$ is the vector of algebraic variables [19]. We restrict our attention to the following semi-explicit DAEs [20]. The DAEs read: find $\boldsymbol{\theta}(t)$ and $\boldsymbol{\gamma}(t)$ such that:

$$\boldsymbol{\theta}(0) = \mathbf{0} \tag{1}$$

$$\dot{\boldsymbol{\theta}}(t, \boldsymbol{\mu}) = \boldsymbol{\varphi}(\boldsymbol{\theta}, \boldsymbol{\gamma}, \boldsymbol{\mu}), \quad \forall t \in [0, t_f] \tag{2}$$

$$\mathbf{0} = \boldsymbol{\psi}(\boldsymbol{\theta}, \boldsymbol{\gamma}) \tag{3}$$

where $t_f$ is a large time instant such that the time interval contains all transient effects submitted to machine learning step. If the given initial values for the differential variables are nonzero, then $\boldsymbol{\theta}$ is the differential variable vector minus its initial value. We assume that there is a bijection between each component of the BG and the row indexes of $\boldsymbol{\varphi} \in \mathbb{R}^{\mathcal{N}_\theta}$ and $\boldsymbol{\psi} \in \mathbb{R}^{\mathcal{N}_\gamma}$. Therefore, we can define an index for each component of the BG, according to the row indexes in $\boldsymbol{\varphi}$ and the row indexes in $\boldsymbol{\psi}$ shifted by $\mathcal{N}_\theta$. By considering parametric DAEs, several simulation data can be produced by solutions of the DAE system for various values of $\boldsymbol{\mu}$ in a training set $\mathcal{D}$.

## 2.2 Hybrid modeling incorporating machine learning and a reduced bond graph

The variables of the original Bond Graph are split into a set of primary variables, a set of secondary variables and a set of tertiary variables, by following an unsupervised machine learning step. The primary variables indexes sets are denoted by $\mathcal{P}^\theta$ and $\mathcal{P}^\gamma$ respectively for variables related to $\boldsymbol{\theta}$ and $\boldsymbol{\gamma}$. The secondary variables indexes sets are denoted by $\mathcal{S}^\theta$ and $\mathcal{S}^\gamma$. $\mathcal{T}^\theta$ and $\mathcal{T}^\gamma$ are the sets of indexes for tertiary variables. Each variable is either primary or secondary or tertiary, such that the following property holds:

$$\mathcal{P}^\theta \cup \mathcal{S}^\theta \cup \mathcal{T}^\theta = \{1, \dots, \mathcal{N}_\theta\} \tag{4}$$

$$\mathcal{P}^\gamma \cup \mathcal{S}^\gamma \cup \mathcal{T}^\gamma = \{1, \dots, \mathcal{N}_\gamma\} \tag{5}$$

The primary differential variables are $\boldsymbol{\theta}[\mathcal{P}^\theta]$, where $[\mathcal{P}^\theta]$ denotes a restriction of a vector to rows whose indexes are in $\mathcal{P}^\theta$. Similarly, we define the primary algebraic variables as $\boldsymbol{\gamma}[\mathcal{P}^\gamma]$. In the sequel, without loss of generality, we assume that:

$$\mathcal{P}^{(x)} = \{1, \dots, card(\mathcal{P}^{(x)})\} \tag{6}$$

$$\{1, \dots, \mathcal{N}_x\} = \{\mathcal{P}^{(x)}, \mathcal{S}^{(x)}, \mathcal{T}^{(x)}\} \tag{7}$$

with $x = \boldsymbol{\theta}$ or $x = \boldsymbol{\gamma}$.

The RBG contains all the primary variables, but none of the tertiary variables. More precisely, the components of the RBG are the components whose indexes are in the set $\mathcal{P}^\theta \cup (\mathcal{N}_\theta + \mathcal{P}^\gamma)$, where $(\mathcal{N}_\theta + \mathcal{P}^\gamma)$ means that indexes of the set $\mathcal{P}^\gamma$ are shifted by $\mathcal{N}_\theta$. Then the DAEs of the RBG are:

$$\boldsymbol{\theta}(0)[\mathcal{P}^\theta] = \mathbf{0} \tag{8}$$

$$\dot{\boldsymbol{\theta}}(t, \boldsymbol{\mu})[\mathcal{P}^\theta] = \boldsymbol{\varphi}(\boldsymbol{\theta}, \boldsymbol{\gamma}, \boldsymbol{\mu})[\mathcal{P}^\theta], \quad \forall t \in [0, t_f] \tag{9}$$

$$\mathbf{0} = \boldsymbol{\psi}(\boldsymbol{\theta}, \boldsymbol{\gamma})[\mathcal{P}^\gamma] \tag{10}$$

We assume that, for given $t$, $\boldsymbol{\theta}$, $\boldsymbol{\gamma}[\mathcal{S}^\gamma]$, $\boldsymbol{\gamma}[\mathcal{T}^\gamma]$ and $\boldsymbol{\mu}$, the solution $\boldsymbol{\gamma}[\mathcal{P}^\gamma]$ of $\boldsymbol{\psi}(t, \boldsymbol{\theta}, \boldsymbol{\gamma})[\mathcal{P}^\gamma] = \mathbf{0}$ is always unique, where:

$$\boldsymbol{\gamma} = \begin{bmatrix} \boldsymbol{\gamma}[\mathcal{P}^\gamma] \\ \boldsymbol{\gamma}[\mathcal{S}^\gamma] \\ \boldsymbol{\gamma}[\mathcal{T}^\gamma] \end{bmatrix} \tag{11}$$

In the sequel, the primary algebraic variables are the variables that are directly coupled to the primary differential variables, such that:

$$\mathcal{P}^\gamma = \left\{ i \in \{1, \ldots, \mathcal{N}_\gamma\}, \sum_{j \in \mathcal{P}^\theta} \left|\frac{\partial^2 \psi}{\partial \gamma_i \partial \theta_j}\right| + \sum_{j \in \mathcal{P}^\theta} \left|\frac{\partial \varphi_j}{\partial \gamma_i}\right| > 0 \quad \forall \boldsymbol{\mu} \in \mathcal{D} \right\} \tag{12}$$

The tertiary variables are original BG variables that are not involved in the RBG equations. The following properties hold:

$$\mathcal{T}^\theta = \left\{ i \in \{1, \ldots, \mathcal{N}_\theta\} \backslash \mathcal{P}^\theta, \sum_{j \in \mathcal{P}^\theta} \left|\frac{\partial \varphi_j}{\partial \theta_i}\right| + \sum_{j \in \mathcal{P}^\gamma} \left|\frac{\partial \psi_j}{\partial \theta_i}\right| = 0 \quad \forall \boldsymbol{\mu} \in \mathcal{D} \right\} \tag{13}$$

$$\mathcal{T}^\gamma = \left\{ i \in \{1, \ldots, \mathcal{N}_\gamma\} \backslash \mathcal{P}^\gamma, \sum_{j \in \mathcal{P}^\theta} \left|\frac{\partial \varphi_j}{\partial \gamma_i}\right| + \sum_{j \in \mathcal{P}^\gamma} \left|\frac{\partial \psi_j}{\partial \gamma_i}\right| = 0 \quad \forall \boldsymbol{\mu} \in \mathcal{D} \right\} \tag{14}$$

Then, the complementary sets define the secondary variables:

$$\mathcal{S}^\theta = \{1, \ldots, \mathcal{N}_\theta\} \backslash (\mathcal{P}^\theta \cup \mathcal{T}^\theta) \tag{15}$$

$$\mathcal{S}^\gamma = \{1, \ldots, \mathcal{N}_\gamma\} \backslash (\mathcal{P}^\gamma \cup \mathcal{T}^\gamma) \tag{16}$$

In the sequel, we restrict our attention to algebraic equations such that there are no secondary algebraic variables:

$$\mathcal{S}^\gamma = \emptyset \tag{17}$$

Let's introduce notations for primary and secondary variables:

$$\boldsymbol{\theta}^\mathcal{P} = \boldsymbol{\theta}[\mathcal{P}^\theta] \tag{18}$$

$$\boldsymbol{\theta}^\mathcal{S} = \boldsymbol{\theta}[\mathcal{S}^\theta] \tag{19}$$

$$\boldsymbol{\gamma}^\mathcal{P} = \boldsymbol{\gamma}[\mathcal{P}^\gamma] \tag{20}$$

Since the tertiary variables are not directly coupled to primary variables, they are not variables of the components involved in the RBG. Then, when applying the Bond Graph methodology to the components of the RBG, we get the following DAEs that ignore tertiary variables:

$$\boldsymbol{\theta}^\mathcal{P}(0) = \boldsymbol{0} \tag{21}$$

$$\dot{\boldsymbol{\theta}}^\mathcal{P}(t, \boldsymbol{\mu}) = \boldsymbol{\varphi}^\mathcal{P}(\boldsymbol{\theta}^\mathcal{P}, \boldsymbol{\gamma}^\mathcal{P}, \boldsymbol{\theta}^\mathcal{S}, \boldsymbol{\mu}), \quad \forall t \in [0, t_f] \tag{22}$$

$$\boldsymbol{0} = \boldsymbol{\psi}^\mathcal{P}(\boldsymbol{\theta}^\mathcal{P}, \boldsymbol{\gamma}^\mathcal{P}, \boldsymbol{\theta}^\mathcal{S}) \tag{23}$$

where $\boldsymbol{\varphi}^\mathcal{P}(\boldsymbol{\theta}^\mathcal{P}, \boldsymbol{\gamma}^\mathcal{P}, \boldsymbol{\theta}^\mathcal{S}, \boldsymbol{\mu}) = \boldsymbol{\varphi}(\boldsymbol{\theta}, \boldsymbol{\gamma}, \boldsymbol{\mu})[\mathcal{P}^\theta]$ and $\boldsymbol{\psi}^\mathcal{P}(\boldsymbol{\theta}^\mathcal{P}, \boldsymbol{\gamma}^\mathcal{P}, \boldsymbol{\theta}^\mathcal{S}) = \boldsymbol{\psi}(\boldsymbol{\theta}, \boldsymbol{\gamma})[\mathcal{P}^\gamma]$. Obviously, there is missing closure equations on secondary variables $\boldsymbol{\theta}^\mathcal{S}$. The following artificial neural network (ANN) is proposed to supplement the equations of the RBG:

$$\boldsymbol{\theta}_i^\mathcal{S} = f^{(1)}\left(\sum_{j=1}^{\text{card}(\mathcal{P}^\theta)} w_{ij}^{(1)} \boldsymbol{\theta}_i^\mathcal{P} + b_i^{(1)}\right), \quad i = 1, \ldots, \text{card}(\mathcal{S}^\theta) \tag{24}$$

It is a fully-connected one-layer artificial neural network. This neural network receives the principal differential variables as inputs and predicts the secondary variables as modulation signals for the RBG. Equations (21) to (24) are the governing equations of the proposed hybrid modeling that couples machine learning with a reduced bond graph. In practice, the ANN is inserted as a supplementary component in the RBG. Moreover, a discrete sampling of the time interval is used to setup an explicit time integration scheme. Then the following recurrent hybrid model is considered in practice:

$$\boldsymbol{\theta}^{\mathcal{P}}(0) = \boldsymbol{0} \tag{25}$$

$$\boldsymbol{\theta}^{\mathcal{S}}(0) = \boldsymbol{0} \tag{26}$$

$$\boldsymbol{\theta}^{\mathcal{P}}(t_{n+1}) = \boldsymbol{\theta}^{\mathcal{P}}(t_n) + (t_{n+1} - t_n)\,\boldsymbol{\varphi}^{\mathcal{P}}\big(\boldsymbol{\theta}^{\mathcal{P}}(t_n), \boldsymbol{\gamma}^{\mathcal{P}}(t_n), \boldsymbol{\theta}^{\mathcal{S}}(t_n), \boldsymbol{\mu}(t_n)\big), \quad \forall t \in [0, t_f] \tag{27}$$

$$\boldsymbol{0} = \boldsymbol{\psi}^{\mathcal{P}}\big(\boldsymbol{\theta}^{\mathcal{P}}(t_n), \boldsymbol{\gamma}^{\mathcal{P}}(t_n), \boldsymbol{\theta}^{\mathcal{S}}(t_n)\big) \tag{28}$$

$$\boldsymbol{\theta}_i^{\mathcal{S}}(t_{n+1}) = f^{(1)}\left(\sum_{j=1}^{\text{card}(\mathcal{P}^\theta)} w_{ij}^{(1)} \boldsymbol{\theta}_i^{\mathcal{P}}(t_{n+1}) + b_i^{(1)}\right), \qquad i = 1, \ldots, \text{card}(\mathcal{S}^\theta) \tag{29}$$

where the parameter $\boldsymbol{\mu}$ have been removed for the sake of simplicity.

For the recovery of tertiary variables $\boldsymbol{\theta}^{\mathcal{T}}$, we propose an additional neural network that we call reconstruction ANN:

$$\boldsymbol{\theta}_i^{\mathcal{T}} = f^{(2)}\left(\sum_{j=1}^{\text{card}(\mathcal{P}^\theta)} w_{ij}^{(2)} \boldsymbol{\theta}_i^{\mathcal{P}} + b_i^{(2)}\right), \qquad i = 1, \ldots, \text{card}(\mathcal{T}^\theta) \tag{30}$$

Hence, the method overcomes the sixth limitation mentioned in introduction, by using this additional neural network. This ANN receives the principal variables as inputs from the RBG in order to compute the tertiary variables without any feedback to the physics-based model. The first ANN model is called at each integration time step, while the latter ANN is only needed at printouts.

The methodology, described in this section, allows thereby building a Reduced Bond Graph (RBG) that well balances the complexity and the accuracy, starting from a high fidelity BG model. The smaller $\mathcal{P}^\theta$ is, the faster should be the prediction using the RBG. This RBG consists of two interacting models: a physics-based model and a one layer ANN.

## *2.3 Dimensionality reduction prior the selection of primary variables*

The primary variables are selected by using a dimensionality reduction method. In this work, the dimensionality reduction is performed via the non-centered Principal Component Analysis (PCA), also known as Singular Value Decomposition (SVD). We consider the training set of simulation data $\{\boldsymbol{\theta}(t_0, \boldsymbol{\mu}_1), \ldots, \boldsymbol{\theta}(t_j, \boldsymbol{\mu}_k), \ldots, \boldsymbol{\theta}(t_m, \boldsymbol{\mu}_P)\}$, where $\{t_0, \ldots, t_m\}$ are sampling time instants and $\{\boldsymbol{\mu}_1, \ldots, \boldsymbol{\mu}_P\} \subset \mathcal{D}$ are obtained by a design of experiment. The vectors $\boldsymbol{\theta}(t_j, \boldsymbol{\mu}_k)$ are solutions of the original DAE system. The simulation data are rearranged in a matrix form prior applying the truncated singular value decomposition that extracts a reduced basis denoted by $\boldsymbol{V}$ according to a tolerance $\varepsilon_{tol}$:

$$[\boldsymbol{\theta}(t_0, \boldsymbol{\mu}_1), \ldots, \boldsymbol{\theta}(t_m, \boldsymbol{\mu}_P)] = \boldsymbol{V}\boldsymbol{S}\boldsymbol{H}^T + \boldsymbol{R}, \qquad \|\boldsymbol{R}\| < \varepsilon_{tol}, \qquad \boldsymbol{V}^T \boldsymbol{R} = \boldsymbol{0} \tag{31}$$

where $\|.\|$ is the Frobenius norm. The matrices $\boldsymbol{V}$ and $\boldsymbol{H}$ are orthonormal. $\boldsymbol{S}$ is the diagonal matrix of the first singular values, which are positive and arranged in a decreasing order. The number of columns in $\boldsymbol{V}$ is denoted $N$. Since it is usually smaller than $\mathcal{N}_\theta$, $\boldsymbol{V}$ is termed reduced-basis matrix.

We assume that it exists a discrete sampling of the continuous set $[0, t_f] \times \mathcal{D}$ such that an exact matrix $\overline{\boldsymbol{V}}$ can be computed by the truncated singular value decomposition with $\varepsilon_{tol} = 0$ such that:

$$\|\boldsymbol{\theta}(t, \boldsymbol{\mu}) - \overline{\boldsymbol{V}}\overline{\boldsymbol{V}}^T \boldsymbol{\theta}(t, \boldsymbol{\mu})\| \leq \varepsilon_S \qquad \forall (t, \boldsymbol{\mu}) \in [0, t_f] \times \mathcal{D} \tag{32}$$

where $\varepsilon_S$ is a sampling tolerance. In the sequel, by following a usual protocol in machine learning, the accuracy of the sampling procedure is checked by considering a testing set of parameters denoted by $\{\widehat{\boldsymbol{\mu}}_1, \ldots, \widehat{\boldsymbol{\mu}}_{\widehat{P}}\}$.

### 2.4 Principal variables

By using the reduced matrix $\boldsymbol{V}$, a reduced approximation of $\boldsymbol{\theta}$ can be introduced, when considering the simulation data of the training set. The vector of reduced coordinates of this approximation is denoted $\boldsymbol{g} \in \mathbb{R}^{\mathcal{N}_\theta \times N}$. It can be computed by using the orthogonal projection of the data on the reduced basis:

$$\boldsymbol{g}(t_j, \boldsymbol{\mu}_k) = \boldsymbol{V}^T \boldsymbol{\theta}(t_j, \boldsymbol{\mu}_k) \tag{33}$$

By construction the following property holds:

$$\sum_{j=0}^{m} \sum_{k=1}^{P} \|\boldsymbol{\theta}(t_j, \boldsymbol{\mu}_k) - \boldsymbol{V} \boldsymbol{g}(t_j, \boldsymbol{\mu}_k)\|^2 \leq \varepsilon_{tol} \tag{34}$$

Proof:

$$[\boldsymbol{V}^T \boldsymbol{\theta}(t_0, \boldsymbol{\mu}_1), \ldots, \boldsymbol{V}^T \boldsymbol{\theta}(t_m, \boldsymbol{\mu}_P)] = \boldsymbol{S} \boldsymbol{H}^T \tag{35}$$

$$\sum_{j=0}^{m} \sum_{k=1}^{P} \|\boldsymbol{\theta}(t_j, \boldsymbol{\mu}_k) - \boldsymbol{V} \boldsymbol{g}(t_j, \boldsymbol{\mu}_k)\|^2 = \|\boldsymbol{R}\|^2 \tag{36}$$

Thanks to the recent work on empirical interpolation of reduced bases [21], the reduced coordinates $\boldsymbol{g}$ related to the training set can also be computed by using $N$ interpolation points, or interpolation indexes, related to the column vectors of $\boldsymbol{V}$. Here, $\mathcal{P}^\theta$ is the set of these interpolation indexes such that:

$$\boldsymbol{g}(t_j, \boldsymbol{\mu}_k) = \left(\boldsymbol{V}[\mathcal{P}^\theta, :]\right)^{-1} \boldsymbol{\theta}(t_j, \boldsymbol{\mu}_k)[\mathcal{P}^\theta] \tag{37}$$

where the notation $[\mathcal{P}^\theta, :]$ means that we consider a selection of rows, in a matrix, whose indexes are in the set $\mathcal{P}^\theta$.

In the proposed modeling approach, we state that the feature extraction performed by the non-centered PCA is a selection of principal physical variables.

### 2.5 Theoretical results and stabilization of the hybrid model

In order to show theoretical results, in this section, we restrict our attention to linear ANNs by choosing linear activation functions $f^{(1)}(x) = x$ and $f^{(2)}(x) = x$.
Then, if $\boldsymbol{V}$ is an exact reduced basis, the best linear ANNs are such that:

$$\boldsymbol{W}^{(1)} = \boldsymbol{V}[\mathcal{S}^\theta, :] \left(\boldsymbol{V}[\mathcal{P}^\theta, :]\right)^{-1}, \quad \boldsymbol{b}^{(1)} = \boldsymbol{0} \tag{38}$$

$$\boldsymbol{W}^{(2)} = \boldsymbol{V}[\mathcal{T}^\theta, :] \left(\boldsymbol{V}[\mathcal{P}^\theta, :]\right)^{-1}, \quad \boldsymbol{b}^{(2)} = \boldsymbol{0} \tag{39}$$

Here, an exact reduced basis means that for all simulation data $\boldsymbol{\theta}^*$ in the validation set the following properties hold:

$$\boldsymbol{\theta}^* = \boldsymbol{V} \left(\boldsymbol{V}[\mathcal{P}^\theta, :]\right)^{-1} \boldsymbol{\theta}^*[\mathcal{P}^\theta] \tag{40}$$

$$\boldsymbol{\theta}^*[\mathcal{S}^\theta] = \boldsymbol{W}^{(1)} \boldsymbol{\theta}^*[\mathcal{P}^\theta] \tag{41}$$

$$\boldsymbol{\theta}^*[\mathcal{T}^\theta] = \boldsymbol{W}^{(2)} \boldsymbol{\theta}^*[\mathcal{P}^\theta] \tag{42}$$

Proof: Let's consider simulation data $\boldsymbol{\theta}^*$ and $\boldsymbol{\gamma}^*$ in the validation set. Let's denote by $\boldsymbol{\mu}^*$ the related vector of BG inputs. Let's assume that $\boldsymbol{\theta}^{\mathcal{P}}(t_n) = \boldsymbol{\theta}^*(t_n)[\mathcal{P}^\theta]$ and $\boldsymbol{\theta}^{\mathcal{S}}(t_n) = \boldsymbol{\theta}^*(t_n)[\mathcal{S}^\theta]$, with $\boldsymbol{\mu} = \boldsymbol{\mu}^*$. Then, $\boldsymbol{\psi}^{\mathcal{P}}(\boldsymbol{\theta}^*(t_n)[\mathcal{P}^\theta], \boldsymbol{\gamma}^{\mathcal{P}}(t_n), \boldsymbol{\theta}^*(t_n)[\mathcal{S}^\theta]) = \boldsymbol{0}$, therefore $\boldsymbol{\gamma}^{\mathcal{P}}(t_n) = \boldsymbol{\gamma}^*(t_n)[\mathcal{P}^\gamma]$. Hence, $\boldsymbol{\theta}^{\mathcal{P}}(t_{n+1}) = \boldsymbol{\theta}^*(t_{n+1})[\mathcal{P}^\theta]$, so $\boldsymbol{\theta}^{\mathcal{S}}(t_{n+1}) = \boldsymbol{W}^{(1)} \boldsymbol{\theta}^*(t_{n+1})[\mathcal{P}^\theta] = \boldsymbol{\theta}^*(t_{n+1})[\mathcal{S}^\theta]$. Since the initial values are exact, the proposed hybrid model propagates the exact solution $\boldsymbol{\theta}^*[\mathcal{P}^\theta]$ and $\boldsymbol{\theta}^*[\mathcal{S}^\theta]$. Therefore the reconstruction ANN recovers exactly $\boldsymbol{\theta}^*[\mathcal{T}^\theta]$ via $\boldsymbol{\theta}^{\mathcal{T}} = \boldsymbol{W}^{(2)} \boldsymbol{\theta}^*[\mathcal{P}^\theta]$. Since the reduced approximation is computed via the singular value decomposition of the available training set simulation data, it is the best low rank approximation of these data according to the Young Eckart theorem. Then, $\boldsymbol{W}^{(1)}, \boldsymbol{b}^{(1)}, \boldsymbol{W}^{(2)}$ and $\boldsymbol{b}^{(2)}$, are the best meta-parameters for linear activation function in the proposed hybrid modeling.

For practical reasons of stability, we found that we should restrict the number of modes involved in the ANNs. We restrict them to $\widetilde{N}$ modes. Then the projection of the simulation data on the reduced basis reads:

$$\widetilde{\boldsymbol{g}}(t_j, \boldsymbol{\mu}_k) = \underset{\boldsymbol{g}^*}{\mathrm{argmin}} \| \boldsymbol{V}[\mathcal{P}^\theta, 1:\widetilde{N}] \, \boldsymbol{g}^* - \boldsymbol{\theta}(t_j, \boldsymbol{\mu}_k)[\mathcal{P}^\theta]\| \quad \forall j, k \tag{43}$$

Then the reduced coordinates read:

$$\widetilde{\boldsymbol{g}}(t_j, \boldsymbol{\mu}_k) = \left(\boldsymbol{V}[\mathcal{P}^\theta, 1:\widetilde{N}]^T \boldsymbol{V}[\mathcal{P}^\theta, 1:\widetilde{N}]\right)^{-1} \boldsymbol{V}[\mathcal{P}^\theta, 1:\widetilde{N}]^T \boldsymbol{\theta}(t_j, \boldsymbol{\mu}_k)[\mathcal{P}^\theta] \tag{44}$$

So, the weights of the stabilized ANNs are:

$$\boldsymbol{W}^{(1)} = \boldsymbol{V}[\mathcal{S}^\theta, 1:\widetilde{N}] \left(\boldsymbol{V}[\mathcal{P}^\theta, 1:\widetilde{N}]^T \boldsymbol{V}[\mathcal{P}^\theta, 1:\widetilde{N}]\right)^{-1} \boldsymbol{V}[\mathcal{P}^\theta, 1:\widetilde{N}]^T, \quad \boldsymbol{b}^{(1)} = \boldsymbol{0} \tag{45}$$

$$\boldsymbol{W}^{(2)} = \boldsymbol{V}[\mathcal{T}^\theta, 1:\widetilde{N}] \left(\boldsymbol{V}[\mathcal{P}^\theta, 1:\widetilde{N}]^T \boldsymbol{V}[\mathcal{P}^\theta, 1:\widetilde{N}]\right)^{-1} \boldsymbol{V}[\mathcal{P}^\theta, 1:\widetilde{N}]^T, \quad \boldsymbol{b}^{(2)} = \boldsymbol{0} \tag{46}$$

## 3   Illustrative example

### 3.1   Illustrative original bond graph model

In Bond Graph approach applied to thermal domain, two elements exchanging energy are linked via a line (Bond). The heat flux $\dot{Q}$ exchanged between the two elements is expressed as the product of the temperature $T$ (effort variable) and the entropy flux $\dot{S}$ (flow variable): $\dot{Q} = T.\dot{S}$. The line is also completed with a half-headed arrow indicating the positive direction of heat transfer, and a causality stroke indicating which of the two elements receives the effort variable and returns the flow one as shown in the left image of Figure 1. Since the entropy flux $\dot{S}$ is difficult to measure, it is more common in thermal engineering to use the heat flux $\dot{Q}$ as a flow variable instead of $\dot{S}$. The modeling approach that uses the temperature $T$ as effort variable and the heat flux $\dot{Q}$ as flow variable is called Pseudo Bond Graph (PBG) modeling approach [22]. In this approach, the product of the effort variable and the flow variable is no longer a power, but the global characteristics of a BG in terms of structure, causality and equations are conserved in PBGs. The PBG modeling approach applied to thermal domain is shown in the right image of Figure 1.

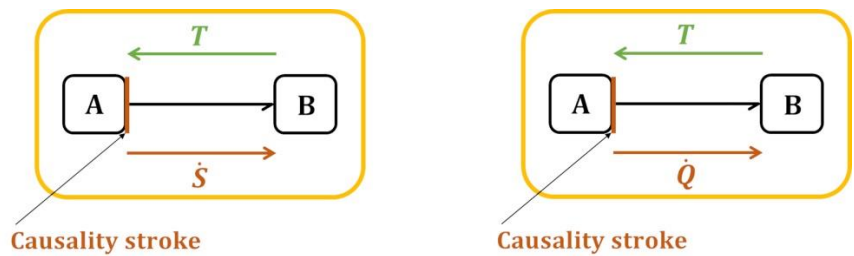

**Figure 1.** *Left*: Bond Graph modeling. *Right*: Pseudo-Bond Graph modeling.

As part of this work, we will build the cabin thermal models using the AMESim software (LMS Imagine.Lab) which is based on the PBG approach. We note that this software does not represent neither the causality stroke nor the half-headed arrow. Then, in order to make the bond graphs more understandable, we indicate on the sketches the positive direction of heat fluxes. In the sequel, we refer to pseudo bond graph models in abbreviate manner as bond graph models.

As an illustrative example to the hybrid modeling methodology proposed in this paper, we consider a simplified cabin system composed of two cabin walls (a roof and a windshield) and one air zone. The two walls are connected on their internal sides to the unique air zone, and on their external sides to the ambient air. The energetic exchanges phenomena that we consider are the conduction through the walls, the internal convection between the internal sides of the walls and the air zone, and finally the external convection between the external sides of the walls and the ambient air.

Each cabin wall is characterized by its total mass $m_x$, its thermal conductivity $\lambda_x$, its specific heat capacity $C_{p_x}$, its thickness $E_x$ and its surface $S_x$; where $x = r$ for the roof wall and $x = w$ for the windshield wall. In this illustrative example, we consider the following values:

| Parameter | Notation (unit) | Windshield ($x = w$) | Roof ($x = r$) |
|---|---|---|---|
| Total mass | $m_x$ $(kg)$ | 14.8525 | 49.708 |
| Surface | $S_x$ $(m^2)$ | 1.3 | 3.4 |
| Thickness | $E_x$ $(mm)$ | 5 | 20 |
| Specific heat capacity | $C_{p_x}$ $(J.kg^{-1}.K^{-1})$ | 829 | 814.5 |
| Thermal conductivity | $\lambda_x$ $(W.m^{-1}.K^{-1})$ | 0.55 | 0.042 |

**Table 1.** Cabin walls characteristics for the illustrative example.

The sketch of the illustrative original bond graph is shown in the Figure 2.

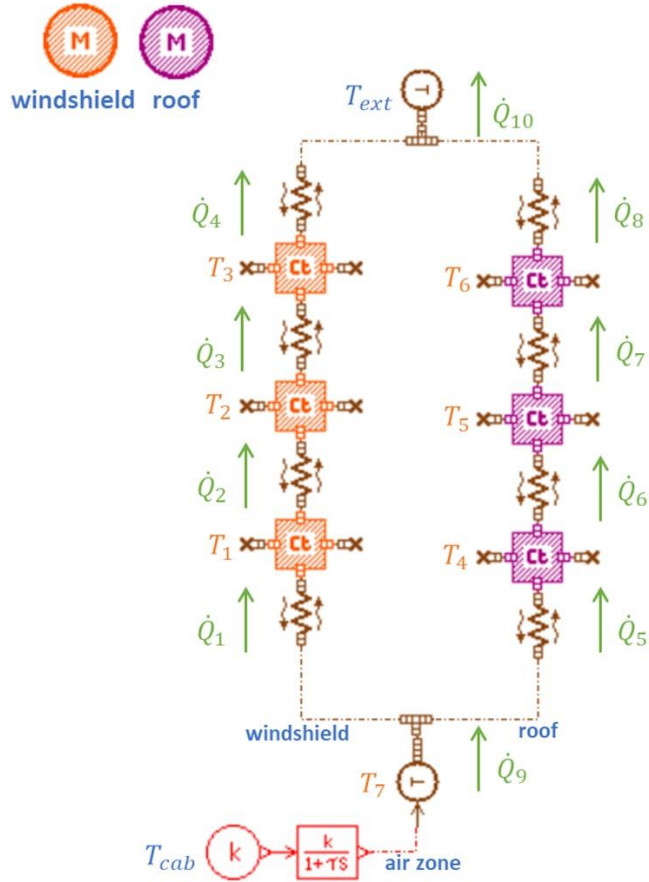

**Figure 2.** Illustrative original Bond Graph.

$T_1$, $T_2$ and $T_3$ are respectively the internal, the middle and the external temperature of the windshield. $T_4$, $T_5$ and $T_6$ are respectively the internal, the middle and the external temperature of the roof. $T_7 = T_{air}$ is the air zone temperature, $T_{ext}$ is the ambient temperature and $T_{cab}$ is the reference comfort temperature set by the driver. $\dot{Q}_1$ (respectively $\dot{Q}_5$) is the internal convective heat flux from the air zone to the windshield (respectively to the roof). $\dot{Q}_4$ (respectively $\dot{Q}_8$) is the external convective heat flux from the air zone to the windshield (respectively to the roof). $\dot{Q}_2$ and $\dot{Q}_3$ are respectively the conduction heat fluxes from the internal side to the middle level and the conduction heat flux from the middle level to the external side of the windshield. $\dot{Q}_6$ and $\dot{Q}_7$ are respectively the conduction heat fluxes from the internal side to the middle level and the conduction heat flux from the middle level to the external side of the roof. $\dot{Q}_9$ (respectively $\dot{Q}_{10}$) is the total internal convective heat flux from the air zone to the internal sides (respectively to the external) of the walls.

We use AMESim's thermal library to model the cabin walls and we take positive the heat fluxes directed from the internal side of the cabin to its external side. To model the air zone, we use a first order lag from the AMESim's signal library. The BG model shown in Figure 2 solves the following linear DAEs:

$$m_w \cdot C_{p_w} \cdot \frac{dT_i}{dt} = \dot{Q}_i - \dot{Q}_{i+1}, \qquad i = 1, \ldots, 3 \tag{47}$$

$$m_r \cdot C_{p_r} \cdot \frac{dT_i}{dt} = \dot{Q}_{i+1} - \dot{Q}_{i+2}, \qquad i = 4, \ldots, 6 \tag{48}$$

$$\frac{dT_7}{dt} = \frac{T_{cab} - T_7}{\tau} \tag{49}$$

$$\dot{Q}_1 = h^{int} S_w (T_7 - T_1) \tag{50}$$

$$\dot{Q}_i = \frac{2 \cdot \lambda_w \cdot S_w}{E_w}(T_{i-1} - T_i), \qquad i = 2, 3 \tag{51}$$

$$\dot{Q}_4 = h^{ext} S_w (T_3 - T_{ext}) \tag{52}$$

$$\dot{Q}_5 = h^{int} S_r (T_7 - T_4) \tag{53}$$

$$\dot{Q}_i = \frac{2 \cdot \lambda_r \cdot S_r}{E_r}(T_{i-2} - T_{i-1}), \qquad i = 6, 7 \tag{54}$$

$$\dot{Q}_8 = h^{ext} S_r (T_6 - T_{ext}) \tag{55}$$

$$\dot{Q}_9 = \dot{Q}_1 + \dot{Q}_5 \tag{56}$$

$$\dot{Q}_{10} = \dot{Q}_4 + \dot{Q}_8 \tag{57}$$

where $h^{int}$ and $h^{ext}$ are respectively the internal and the external convective heat transfer coefficients. And $\tau$ is the time constant of the first order lag.

The differential equations (47) and (48) are solved in the thermal capacitance BG models. The differential equation (49) is solved in the first order lag BG model. The algebraic equations (50) to (55) are solved in the thermal conductance BG models. And the algebraic equations (56) and (57) are solved in 0-junction BG models.

For this illustrative BG model, the vector $\boldsymbol{\theta}$ of differential variables and the vector $\boldsymbol{\gamma}$ of algebraic variables are given by:

$$\boldsymbol{\theta} = [T_1, T_2, \ldots, T_7]^T \in \mathbb{R}^{\mathcal{N}_\theta} \qquad \text{with} \quad \mathcal{N}_\theta = 7 \tag{58}$$

$$\boldsymbol{\gamma} = [\dot{Q}_1, \dot{Q}_2, \ldots, \dot{Q}_{10}]^T \in \mathbb{R}^{\mathcal{N}_\gamma} \qquad \text{with} \quad \mathcal{N}_\gamma = 10 \tag{59}$$

### 3.2 Illustrative reduced bond graph model

In this subsection, we build a RBG using two training values of the external convective heat transfer coefficient $\{h_1^{ext}, h_2^{ext}\}$, then test the RBG on a third value $h_3^{ext}$. The values of $h_i^{ext}$ are given in the Table 2.

| $i$ | 1 | 2 | 3 |
|---|---|---|---|
| $h_i^{ext}$ (W/m²/K) | 35 | 10 | 20 |

**Table 2.** The $h^{ext}$ values considered to train and test the illustrative RBG.

Excepting the parameter $h^{ext}$, we keep constant all the other BG inputs (parameters, excitations and initial conditions) as well as the simulations duration and print interval at the values shown in the Table 3:

| Variable | Notation (unit) | Value |
|---|---|---|
| Internal convective heat transfer coefficient | $h^{int}\ (W.m^{-2}.K^{-1})$ | 20 |
| Time constant of the first order lag | $\tau\ (s)$ | 60 |
| Ambient temperature | $T_{ext}\ (°C)$ | -18 |
| Reference control temperature | $T_{cab}\ (°C)$ | 20 |
| Initial states | $\theta_i(0)\ (°C)$ | $T_{ext}$ |
| Simulation duration | $t_f\ (s)$ | 3600 |
| Print interval | $\Delta t\ (s)$ | 1 |

**Table 3.** List of constant BG inputs and simulations setups.

In the training step, we consider a snapshots matrix $A \in \mathbb{R}^{\mathcal{N}_\theta \times (P.m)}$ where $P = 2$ and $m = t_f/\Delta t = 3600$. This snapshots matrix is defined by:

$$A_{ij} = \theta_i(t_k, \boldsymbol{\mu}_p) - \theta_i(0)$$
$$\text{with}\quad j = (p-1)m + k, \quad 1 \le k \le m, \quad 1 \le p \le P \tag{60}$$

where $\boldsymbol{\mu}_p$ is the BG input vector composed of the $h_p^{ext}$ value and all the other constant BG inputs given in the Table 1 and the Table 3.

By applying a SVD to the matrix $A$, we extract $\mathcal{N}_\theta$ singular values that we present in the following table:

| $i$ | $s_i$ | $s_i/s_1$ |
|---|---|---|
| 1 | 5177.9 | 1.0000 |
| 2 | 664.1 | 0.1283 |
| 3 | 389.5 | 0.0752 |
| 4 | 153.7 | 0.0297 |
| 5 | 28.3 | 0.0055 |
| 6 | 5.3 | 0.0010 |
| 7 | 0.7 | 0.0001 |

**Table 4.** List of singular values.

We also extract $N$ empirical modes that form a reduced basis $V \in \mathbb{R}^{\mathcal{N}_\theta \times N}$ where $N = 4$. The reduced basis is given below:

$$V = \begin{bmatrix} -0.3290 & -0.3630 & -0.1841 & -0.0527 \\ -0.3044 & -0.4360 & -0.1997 & -0.0472 \\ -0.2807 & -0.4974 & -0.2457 & -0.0080 \\ -0.5314 & 0.1738 & 0.4238 & -0.6702 \\ -0.2525 & -0.1760 & 0.7418 & 0.5334 \\ -0.0492 & -0.1852 & 0.0350 & 0.3274 \\ -0.6097 & 0.5790 & -0.3672 & 0.3925 \end{bmatrix} \tag{61}$$

By applying the discrete empirical interpolation method (DEIM) algorithm [23] to the reduced basis, we extract $N$ primary differential variables indexed by $\mathcal{P}^\theta$ such that:

$$\mathcal{P}^\theta = \{3, 4, 5, 7\} \tag{62}$$

The primary algebraic variables list $\mathcal{P}^\gamma$ includes all the algebraic variables that are linked to a primary differential variable through an algebraic equation:

$$\mathcal{P}^{\gamma} = \{3, 4, 5, 6, 7\} \qquad (63)$$

We point out that the list $\mathcal{P}^{\gamma}$ does not include the index $\{1\}$ since the air zone temperature $T_7$ is set as a signal source, which means that it can not be influenced by any thermal heat flux (see equation (49)). The tertiary differential variables list $\mathcal{T}^{\theta}$ includes all the differential variables that are not linked to a primary differential variable through a differential equation:

$$\mathcal{T}^{\theta} = \{1\} \qquad (64)$$

The list $\mathcal{T}^{\gamma}$ is the complementary of $\mathcal{P}^{\gamma}$ since there are no secondary algebraic variables:

$$\mathcal{T}^{\gamma} = \{1, 2, 8, 9, 10\} \qquad (65)$$

The secondary differential variables list $\mathcal{S}^{\theta}$ includes all the differential variables that are linked to a primary differential variable through a differential equation. The list $\mathcal{S}^{\theta}$ is also given by the equation (15):

$$\mathcal{S}^{\theta} = \{2, 6\} \qquad (66)$$

In order to build the RBG, all the BG components related to $\boldsymbol{\theta}^{\mathcal{T}}$ and $\boldsymbol{\gamma}^{\mathcal{T}}$ need to be removed from the initial sketch. The sketch of the RBG is shown in Figure 3. We use the notation $\tilde{T}_i$ (respectively $\tilde{\dot{Q}}_i$) to refer to the approximation of the variable $T_i$ (respectively $\dot{Q}_i$) by the RBG.

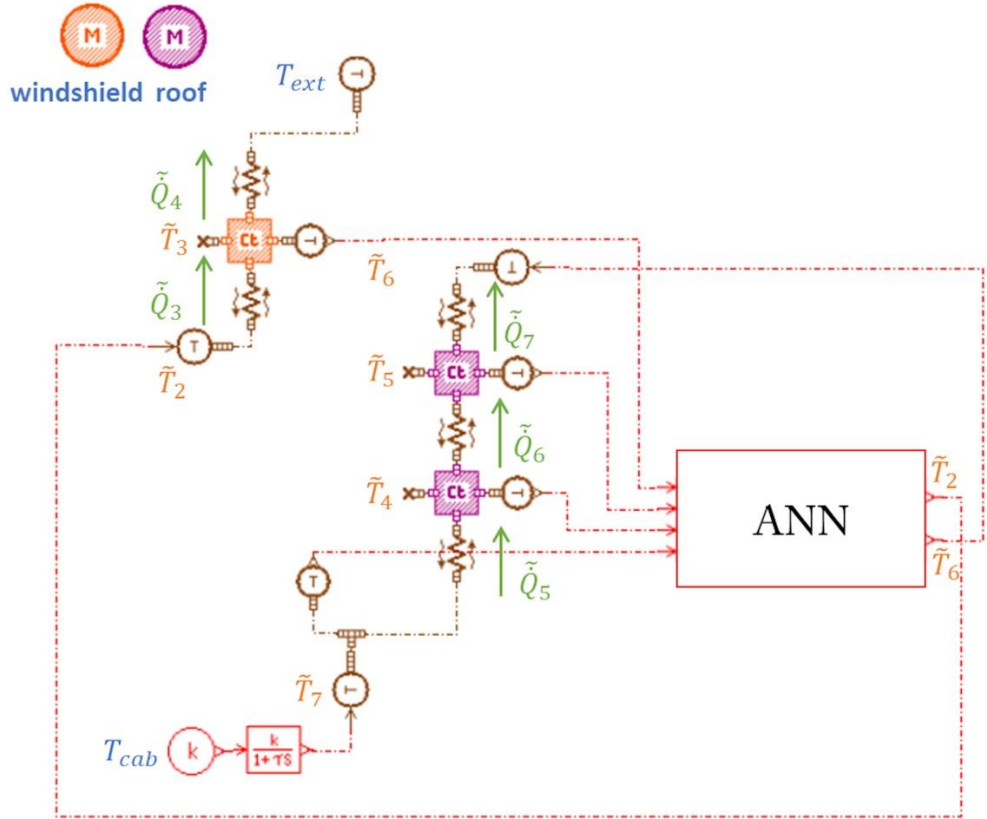

**Figure 3.** Illustrative Reduced Bond Graph.

An ANN is added to the sketch of the RBG. This ANN receives the primary differential variables at its input layer, and computes the secondary differential variables as outputs. More precisely, the ANN computes $\left(\tilde{\boldsymbol{\theta}}^S - \tilde{\boldsymbol{\theta}}^S(0)\right)$ as a function of $\left(\tilde{\boldsymbol{\theta}}^{\mathcal{P}} - \tilde{\boldsymbol{\theta}}^{\mathcal{P}}(0)\right)$ such that:

$$\left(\tilde{\boldsymbol{\theta}}^S - \tilde{\boldsymbol{\theta}}^S(0)\right) = f^{(1)}\left(\boldsymbol{W}^{(1)}\left(\tilde{\boldsymbol{\theta}}^{\mathcal{P}} - \tilde{\boldsymbol{\theta}}^{\mathcal{P}}(0)\right) + \boldsymbol{b}^{(1)}\right) \tag{67}$$

$$\tilde{\boldsymbol{\theta}}^{\mathcal{P}} = [\tilde{T}_3, \tilde{T}_4, \tilde{T}_5, \tilde{T}_7]^T \quad \text{and} \quad \tilde{\boldsymbol{\theta}}^S = [\tilde{T}_2, \tilde{T}_6]^T \tag{68}$$

where $f^{(1)}$ represents two linear activation functions, $\tilde{\boldsymbol{\theta}}^{\mathcal{P}}$ and $\tilde{\boldsymbol{\theta}}^S$ are respectively the approximations of $\boldsymbol{\theta}^{\mathcal{P}}$ and $\boldsymbol{\theta}^S$ by the RBG. And the weights and bias of the ANN are given by:

$$\boldsymbol{W}^{(1)} = \begin{bmatrix} 0.9176 & 0.0699 & 0.0022 & 0.0150 \\ 0.2480 & -0.2331 & 0.2875 & 0.0506 \end{bmatrix}, \qquad \boldsymbol{b}^{(1)} = \begin{bmatrix} 0 \\ 0 \end{bmatrix} \tag{69}$$

If one needs an approximation $\tilde{Q}_9$ of the total internal convective heat flux from the air zone to all walls, then we need an approximation $\tilde{T}_1$ of the tertiary temperature $T_1$. In this case, we use a reconstruction ANN which computes the following supplementary equation at printouts:

$$\left(\tilde{T}_1 - \tilde{T}_1(0)\right) = f^{(2)}\left(\boldsymbol{w}^{(2)}\left(\tilde{\boldsymbol{\theta}}^{\mathcal{P}} - \tilde{\boldsymbol{\theta}}^{\mathcal{P}}(0)\right) + b^{(2)}\right) \tag{70}$$

where $f^{(2)}$ is the linear activation function. And the weights and bias of the reconstruction ANN are given by:

$$\boldsymbol{w}^{(2)} = [0.8385 \quad 0.1051 \quad 0.0002 \quad 0.0619], \qquad b^{(2)} = 0 \tag{71}$$

In order to evaluate the accuracy of the RBG, we compute a Mean Absolute Error (MAE) and a Maximal Absolute Error (MaxAE) as shown in equations (72) and (73). The obtained results are very satisfying in our context.

$$MAE(\boldsymbol{\theta}, \boldsymbol{\mu}_3) = \sum_{i=1}^{\mathcal{N}_{\theta}} \sum_{k=1}^{m} \frac{|\theta_i(t_k, \boldsymbol{\mu}_3) - \tilde{\theta}_i(t_k, \boldsymbol{\mu}_3)|}{\mathcal{N}_{\theta}.m} = 0.11\ °C \tag{72}$$

$$MaxAE(\boldsymbol{\theta}, \boldsymbol{\mu}_3) = \max_{\substack{i=1,\ldots,\mathcal{N}_{\theta} \\ k=1,\ldots,m}} |\theta_i(t_k, \boldsymbol{\mu}_3) - \tilde{\theta}_i(t_k, \boldsymbol{\mu}_3)| = 0.64\ °C \tag{73}$$

## 4 Industrial application

### 4.1 Industrial high fidelity model

In this fourth section, we apply the RBG methodology to the Renault Scenic 3 industrial cabin model which is composed of 126 walls and 30 air zones. We split each wall into an internal wall connected to cabin air, and an external wall connected or not to the ambience. For each cabin wall, we compute then an internal and an external wall temperature using two thermal capacitance models. We consider the following physical phenomena: the conduction through walls, the longwave radiation between internal walls, the longwave radiation between external walls and the ambience, the direct solar shortwave radiation on external walls, the transmitted solar shortwave radiation to internal walls through windows, the internal convection between each internal wall and the connected air zone, the external convection between the ambient air and the external walls that are connected to ambience, the air ventilation inside the cabin which results in humid airflow exchanges between air zones, the air recirculation through the Heating, Ventilating and Air Conditioning system (HVAC), the air extraction from the cabin, and finally the air blown by the HVAC inside the cabin through the air vents.

Unlike the illustrative model of section 3, the industrial model computes air zones temperatures and absolute humidities through energy and mass balances. Besides, radiative heat exchanges are taken into account which makes the industrial cabin model nonlinear. The DAE system solved has the form of the equations (1) to (3), where the functions $\varphi(\theta, \gamma, \mu)$ and $\psi(\theta, \gamma)$ are nonlinear functions. In addition, the vectors $\theta$ and $\gamma$ are given by:

$$\theta = \begin{bmatrix} T^{wi} \\ T^{we} \\ h \\ x \end{bmatrix} \in \mathbb{R}^{\mathcal{N}_\theta} \quad , \quad \gamma = \begin{bmatrix} T^a \\ \dot{Q} \\ r \end{bmatrix} \in \mathbb{R}^{\mathcal{N}_\gamma} \tag{74}$$

where $T^{wi} \in \mathbb{R}^{\mathcal{N}_w}$ and $T^{we} \in \mathbb{R}^{\mathcal{N}_w}$ are respectively the vectors of internal and external walls temperatures such that $\mathcal{N}_w$ is the cabin walls number. $h \in \mathbb{R}^{\mathcal{N}_a}$ and $x \in \mathbb{R}^{\mathcal{N}_a}$ are respectively the vectors of air zones specific enthalpies and absolute humidities such that $\mathcal{N}_a$ is the cabin air zones number. $T^a \in \mathbb{R}^{\mathcal{N}_a}$ and $r \in \mathbb{R}^{\mathcal{N}_a}$ are respectively the vectors of air zones temperatures and air zones relative humidities. Finally, $\dot{Q} \in \mathbb{R}^{\mathcal{N}_Q}$ is a vector of a large number of heat fluxes involved in the high fidelity BG connections.

For this industrial high fidelity model (HFM), we have:

$$\mathcal{N}_\theta = 2(\mathcal{N}_w + \mathcal{N}_a) = 312 \tag{75}$$

### 4.2 Industrial reduced bond graph model

We aim to build a RBG for cooling purposes. Thus, we consider the parametric space of the Table 5 in which the inlet air temperature is always lower than the ambient temperature.

| Variable | Notation (unit) | Minimal value | Maximal value |
|---|---|---|---|
| Vehicle speed | $V^{veh}$ $(km/h)$ | 0 | 130 |
| Ambient temperature | $T_{ext}$ $(°C)$ | 20 | 45 |
| Ambient relative humidity | $r_{ext}$ $(\%)$ | 0 | 80 |
| Solar irradiance | $I$ $(W/m^2)$ | 0 | 1200 |
| Inlet air mass flow rate | $\dot{m}^{inlet}$ $(kg/h)$ | 100 | 600 |
| Inlet air relative humidity | $r^{inlet}$ $(\%)$ | 0 | 100 |
| Inlet air temperature | $T^{inlet}$ $(°C)$ | 2 | 12 |

**Table 5.** The parametric space considered for the industrial application.

In order to capture various dynamics of the cabin, we use a big amount of simulation data during the training step: through a DOE, we generate 500 training points $\{\mu_1, ..., \mu_P\}$ and 500 test points $\{\hat{\mu}_1, ..., \hat{\mu}_{\hat{P}}\}$ over the considered parametric space; and we set all the simulations durations to four hours. We note that the use of Air Conditioning (AC) makes the absolute humidity of inlet air always lower than or equal to the ambient absolute humidity due to water vapor condensation on the evaporator. By applying a constraint, based on this last physical rule, to the generated training and test points, we eliminate nearly 70 points among the 500 points previously generated.

Since the orders of magnitude in $\theta$ are very different, we build a snapshot matrix based on the following homogeneous multiphysics state vector $\hat{\theta}$:

$$\hat{\theta} = \begin{bmatrix} T^{wi} \\ T^{we} \\ h/C_p^{da} \\ x * 1000 \end{bmatrix} \in \mathbb{R}^{\mathcal{N}_\theta} \tag{76}$$

$$C_p^{da} = 10^3 \, J.kg^{-1}.K^{-1} \tag{77}$$

where $(h/C_p^{da})$ means that all the components of the vector $h$ are divided by the dry air specific heat capacity ($C_p^{da}$) in order to make the specific enthalpies equivalent to temperatures. All the components of the vector $x$ are multiplied by 1000 in order to convert the absolute humidities from the unit $(kg_{water\ vapor}/kg_{dry\ air})$ to the $(g_{water\ vapor}/kg_{dry\ air})$ one with the purpose of making the absolute humidities have the same order of magnitude as temperatures.

By applying the RBG method, we select 29 primary differential variables distributed as follows: 20 walls temperatures, 5 air zones enthalpies and 4 air zones absolute humidities. By using the BG connection tables, we determine the secondary and tertiary variables. In practice, we consider the following connection tables: air zone to air zone connection table for airflow exchanges, air zone to internal walls connection table for the internal convection, external walls to ambient air connection table for the external convection and longwave radiation with the ambience, internal wall to internal wall connection table for the longwave radiation, and windows to internal walls connection table for the transmitted solar shortwave radiation.

We term *primary wall* a wall whose internal or external temperature is primary. The Figure 4 shows two perspective views of the high fidelity cabin model at left and two perspective views of the reduced cabin model at right on which only primary walls are represented. For the high fidelity model (HFM) as well as the reduced order model (ROM), we present an opaque view at the top of the Figure 4 and a transparent view at its down in order to make the cabin's interior walls visible.

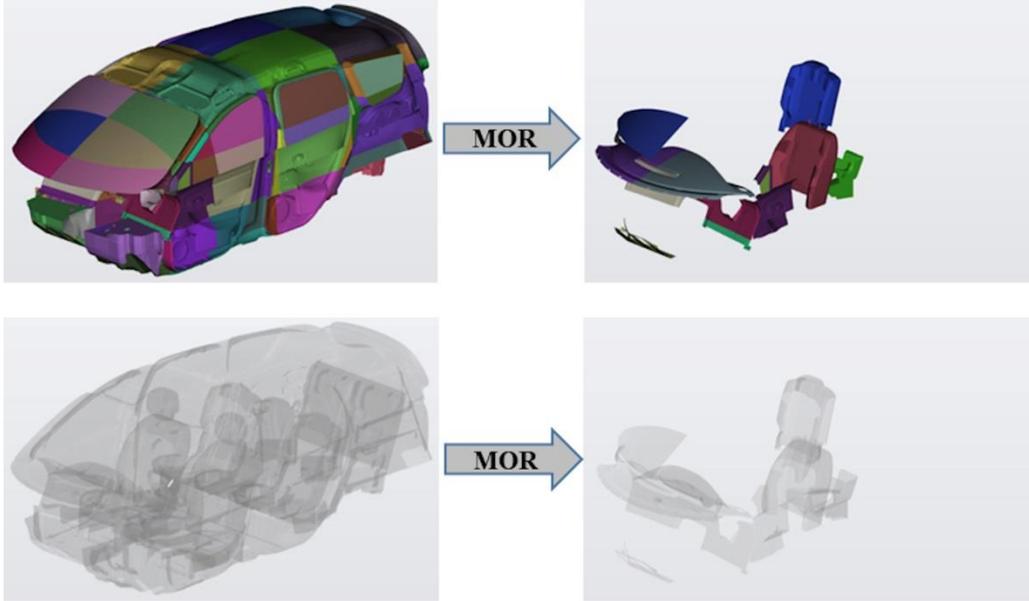

**Figure 4.** *Left top*: opaque view of the HFM. *Left down*: transparent view of the HFM. *Right top*: opaque view of the ROM. *Right down*: transparent view of the ROM.

For this industrial example, we use the Rectified Linear Unit (ReLU) activation function in the ANN to avoid the air zones absolute humidities to be negative.

We evaluate the accuracy of the RBG on three variables of interest ($r^{recy}$, $T^{recy}$ and $T^{head}$) where $r^{recy}$ and $T^{recy}$ are respectively the mean relative humidity and the mean temperature of cabin recirculation air zones, and $T^{head}$ is the driver's head air zone temperature. The variables $r^{recy}$ and $T^{recy}$ are used for HVAC sizing purposes, while the variable $T^{head}$ is used for temperature control applications. We compute a MAE and a MaxAE for each variable $x$ of interest as shown in equations (78) and (79), where $\tilde{x}$ is the approximation of $x$ using the RBG. The Table 6 presents the obtained numerical results. These results are very satisfying in our context.

$$MAE\left(x,\{\hat{\boldsymbol{\mu}}_p\}_{p=1}^{\hat{P}}\right) = \sum_{p=1}^{\hat{P}} \sum_{i=1}^{\mathcal{N}_\theta} \sum_{k=1}^{m} \frac{|x(t_k,\boldsymbol{\mu}_p) - \tilde{x}(t_k,\boldsymbol{\mu}_p)|}{\mathcal{N}_\theta . m . \hat{P}} \tag{78}$$

$$MaxAE\left(x,\{\hat{\boldsymbol{\mu}}_p\}_{p=1}^{\hat{P}}\right) = \max_{\substack{i=1,\ldots,\mathcal{N}_\theta \\ k=1,\ldots,m \\ p=1,\ldots,\hat{P}}} |x(t_k,\boldsymbol{\mu}_p) - \tilde{x}(t_k,\boldsymbol{\mu}_p)| \tag{79}$$

| Variable of interest | MAE | MaxAE |
|---|---|---|
| $T^{head}$ (°C) | 0.06 | 1.53 |
| $T^{recy}$ (°C) | 0.08 | 0.64 |
| $r^{recy}$ (%) | 0.11 | 5.25 |

**Table 6.** Accuracy of the RBG model on the three variables of interest.

We also test the RBG on the validation scenario described by the Table 7. The vehicle speed follows the *WLTC-class-3* homologation driving cycle [24,25]. The cycle's duration is 30 minutes during which the vehicle speed varies between 0 and 131.3 km/h.

| Variable | Notation (unit) | Value |
|---|---|---|
| Vehicle speed | $V^{veh}$ (km/h) | WLTC cycle |
| Ambient temperature | $T_{ext}$ (°C) | 45 |
| Ambient relative humidity | $r_{ext}$ (%) | 40 |
| Solar irradiance | $I$ (W/m²) | 1000 |
| Inlet air mass flow rate | $\dot{m}^{inlet}$ (kg/h) | 450 |
| Inlet air relative humidity | $r^{inlet}$ (%) | 20 |
| Inlet air temperature | $T^{inlet}$ (°C) | 15 |

**Table 7.** BG inputs for the validation scenario.

The external convective power between the cabin external walls and the ambience depends on the vehicle speed. In order to approximate this external convective power using the RBG, we compute all the cabin external walls temperatures using the reconstruction ANN. The Figure 5 plots at left the external convective power from the ambience to external walls and plots at right the mean relative humidity of recirculation air zones using the HFM and the RBG. On this validation scenario, we obtain a simulation speed-up of 10.

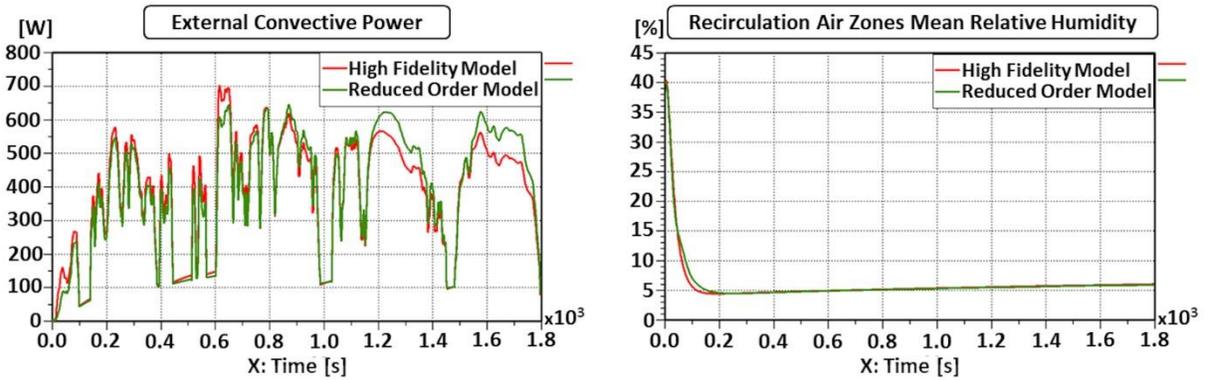

**Figure 5.** The external convective power (*left*) and the mean relative humidity of recirculation air zones (*right*) on the validation scenario by using the HFM and the ROM.

## 5 Conclusion

In this paper, we propose a BG hybrid modeling approach that couples a physics-based RBG with an ANN. The proposed method overcomes the six limitations mentioned in introduction: the method is applicable to nonlinear systems, states meaning preserving, input dependent, applicable to graph models, reduce the BG junction structure and recover all the original variables using a reconstruction ANN. Besides, the methodology is not very intrusive for BGs since one just needs to remove the tertiary BG components from the sketch and add an ANN that supplements the RBG with the secondary variables. In this paper, we propose a MOR-based methodology to calibrate the weights of the two ANNs without using a Backpropagation Through Time. The use of the ReLU activation function avoids the absolute humidities to be negative. In perspective, more complex ANNs containing at least one hidden layer could be considered in order to improve the accuracy of the RBG or its speed-up.

Furthermore, it would be highly interesting to use the suggested modeling methodology for setting up automotive embedded climate control systems which generally have limited computing resources. For this purpose, a reduced cabin thermal model can supplement the HVAC unit with approximated air zones temperatures and humidities in real time. Besides, thanks to the reconstruction ANN, a multi-zone climate control system could be designed with a minimum number of temperature sensors. This could help either the common automotive manufacturers to improve their thermal comfort service, or the luxury automotive industry to perform some cost reductions in order to make luxury cars more affordable.

## Acknowledgements

Funding: This work was supported by the Association Nationale de la Recherche et de la Technologie (ANRT) [grant number CIFRE 2016/0490].

## Declarations of interest

None.